\documentclass{elsart}

\usepackage{amssymb}
\usepackage{graphicx}

\begin{document}

\begin{frontmatter}




\title{Vertex Reconstruction \\
Using a Single Layer Silicon Detector}

\author{ E.Garc\'{\i}a$^5$,     }
\author{ R.S.Hollis$^5$,     }
\author{ A.Olszewski$^3$,     }   
\author{ I.C.Park$^6$,     }
\author{ M.Reuter$^5$,     }
\author{ G.Roland$^4$,     }
\author{ P.Steinberg$^2$,     }
\author{ K.Wo\'{z}niak$^3$,     }
\author{ A.H.Wuosmaa$^{1,}$\corauthref{praddress}  }

\address{
$^1$~Argonne National Laboratory, Argonne, IL 60439-4843, USA\\
$^2$~Brookhaven National Laboratory, Upton, NY 11973-5000, USA\\
$^3$~Institute of Nuclear Physics PAN, Krak\'{o}w, Poland\\
$^4$~Massachusetts Institute of Technology, Cambridge, MA 02139-4307, 
USA\\
$^5$~University of Illinois at Chicago, Chicago, IL 60607-7059, USA\\
$^6$~University of Rochester, Rochester, NY 14627, USA
}
\corauth[praddress]{present address: Physics Department, Western Michigan University, Kalamazoo MI, 49008-5252, USA}

\begin{abstract}

Typical vertex finding algorithms use reconstructed tracks,
registered in a multi-layer detector, which directly point to 
the common point of origin.
A detector with a single layer of silicon sensors registers the passage of
primary particles only in one place. Nevertheless, the information available
from these hits can also be used to estimate the vertex position, when
the geometrical properties of silicon sensors and the measured
ionization energy losses of the particles are fully exploited.
In this paper the algorithm used for this purpose in the PHOBOS experiment 
is described. The vertex reconstruction performance is studied using simulations and
compared with results obtained from real data.
The very large acceptance of a single-layered multiplicity detector permits
vertex reconstruction for low multiplicity events where other
methods, using small acceptance subdetectors, 
fail because of insufficient number of registered primary tracks.
\end{abstract}

\end{frontmatter}

\begin{figure}[bt]
\includegraphics[width=14cm]{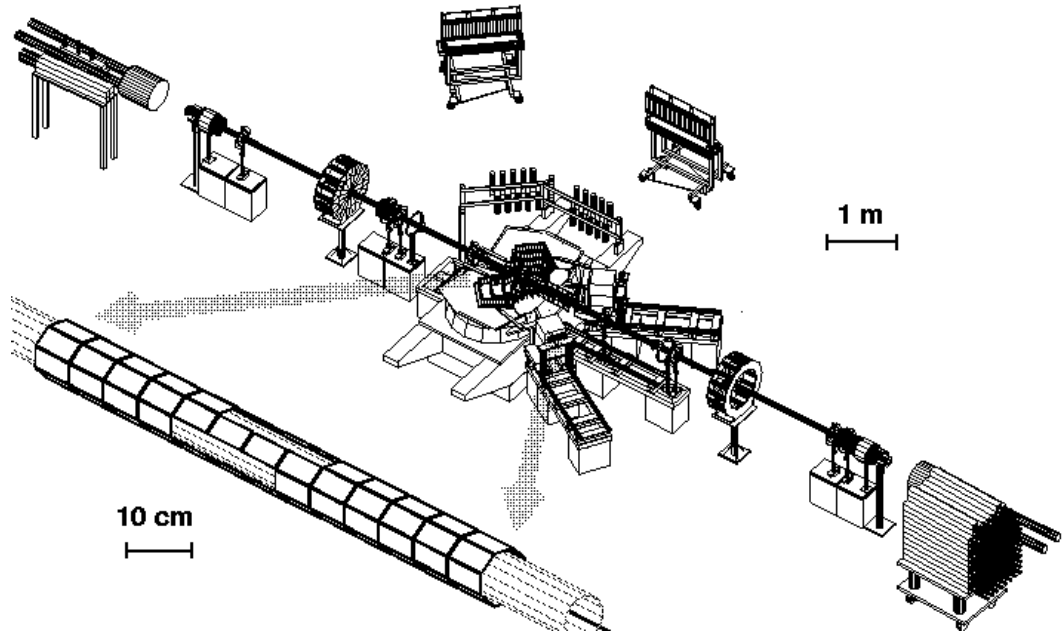}
\caption{ \label{Fig1}
General view of the PHOBOS detector and a magnified drawing 
of the octagonal 
multiplicity detector centered around the nominal interaction point. 
Missing sensors in front side and on top form
the windows for the spectrometer and the vertex detector, respectively.
Identical windows (not visible) are also in the back side and bottom of
this detector. Two beams of ions travel close to the axis of the beam pipe cylinder
(which coincides with the $z$-axis of the PHOBOS coordinate system)
and collide along this axis up to $\pm$1~m from the detector center.
}
\end{figure}

\section{Introduction}
\label{Sec1}

An aspect of the events recorded at the Relativistic Heavy Ion Collider (RHIC) 
critical for all subsequent data analyzes is the position 
of the collision vertex of high energy protons or nuclei. 
In RHIC experiments, the vertex positions are distributed
along the beam axis in a range as large as 2~m about the nominal
center of the apparatus, with a transverse spread smaller than 1~mm.  
The PHOBOS detector \cite{detector} includes several subsystems 
which are used also for determinations of the vertex position: 
the trigger system, the two arm multi-layer spectrometer for measuring charged particles 
trajectories, the vertex detector consisting of two
layers of silicon sensors and the multiplicity detector 
measuring charged particles. Most of the active 
elements of the PHOBOS detector employ silicon sensors 
described in detail in \cite{sensors}. 

The standard vertex reconstruction algorithms in PHOBOS
enable either a full 3-dimensional reconstruction of the vertex using 
the information from the spectrometer, or a more precise determination of
two coordinates (along the beam and in the vertical direction) 
with the help of the vertex detector \cite{time05}. 
However, the range of reconstructed vertex positions along
the beam axis, $z$, for which these detectors are effective, 
corresponding to $-$50~cm to +10~cm for the spectrometer, 
and $-$15~cm to +15~cm for the vertex detector, does not always include 
all collision points. These two subdetectors
have very limited geometrical acceptances (1\% and 5\% for spectrometer and
vertex detector, respectively) and they register very few primary particles
in the events with low multiplicities. The efficiency of vertex reconstruction 
in d+Au and especially in p+p collisions is thus insufficient \cite{time05}.
The octagonal multiplicity detector, ($octagon$), that surrounds the
beam pipe and extends $\pm$55~cm from the nominal interaction point,
covers a much larger geometrical acceptance of 50\% (see Fig.~1). This detector
consists of a single layer of silicon sensors 
arranged parallel to the beam, 
which record the energy loss of charged particles.
As we show in this paper a single hit
from a primary particle provides enough information for a crude estimation
of the vertex position along the beam line. 
With a larger number of such hits the vertex 
position can be calculated with an uncertainty of the order of 1~cm.
The very large acceptance of the $octagon$ detector makes this method very efficient even for 
events with small multiplicities.

\begin{figure}[bt]
\includegraphics[width=7.5cm]{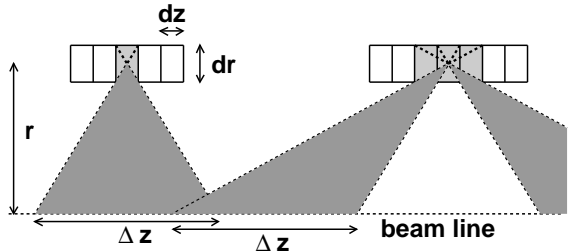}
\caption{ \label{Fig2}
Examples of hits from primary
particles leaving signals in one (left) or three (right)
active elements of the silicon detector and the ranges of possible 
vertex positions, $\Delta z$, defined by simple geometrical calculations.  
}
\end{figure}

\section{Estimation of the vertex position from a single hit}
\label{Sec2}

Silicon sensors of the $octagon$, placed along  the beam line, register primary particles
emitted and thus entering the silicon at many different angles. 
The properties of the hits depend on emission angle and thus on the distance of the hit
from the primary vertex, as shown in Fig.~2. 
A signal in an isolated pad can be left only by a particle 
emerging from a collision that occurred within a range of $z$ given by 
$\Delta z$ = 2\,$r$\,$dz$\,/\,$dr$, where $r$ is the transverse distance from the beam axis to
the pad, $dz$ is the length of the pad in $z$, and $dr$ is the pad
thickness.
For the multiple-pad hits, left by particles emitted at smaller angles,
there exist two ranges of possible vertex positions. A signal in a single pad gives 
an unambiguous vertex position estimate with an error of   
approximately $\sigma_{hit}$~=~$\frac{\Delta z}{2 \sqrt{6}}$. For multiple-pad hits a
similar estimate is possible only if we can select the correct vertex range.
The sensors of the  $octagon$ are placed close to beam, at  $r$=4.5~cm, 
but the pad dimensions, $dz$=0.27~cm and $dr$=0.033~cm, are such that the geometrical 
approach is not precise enough ($\sigma_{hit}$=15~cm). 

\begin{figure}[bt]
\includegraphics[width=7.5cm]{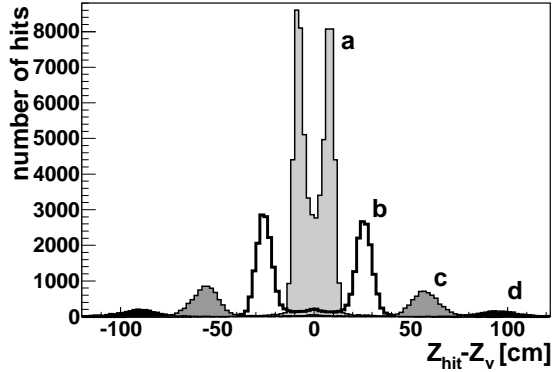}

\caption{ \label{Fig3}
The distribution of the number of hits as a function of the distance of the hit
from the vertex, $Z_{hit}-Z_{v}$, for several values of the energy loss, 
$\Delta E_{norm}$ (2.5, 6, 12 and 25 for histograms 
labeled a, b, c and d, respectively).
}
\end{figure}

The silicon sensors used in the PHOBOS detector register not only the passage 
of the charged particles but also the deposited energy, which 
is proportional to the length of the trajectory in the silicon.
This track length in the $octagon$
silicon sensors depends on the
particle emission angle, and thus the distance of the hit from the
collision vertex is correlated with the deposited energy.
In Fig.~3 we present several histograms of $Z_{hit}-Z_{v}$ 
for hits left by primary particles, obtained 
for several fixed values of the deposited energy loss. They were
obtained using GEANT  \cite{geant}
simulations of the real detector, with accurate calculations of ionization energy losses,
but in which effects of electronics noise, 
energy sharing between pads, digitization and merging of signals
from different particles were not included. 
In order to correct for varying thickness of the silicon sensors 
we are using the normalized energy loss
$\Delta E_{norm}$=$\Delta E$/$\Delta E_{MIP}$, where $\Delta E$ is the actual
deposited energy and $\Delta E_{MIP}$ is the
energy loss for a minimum ionizing particle at normal incidence. 
The hit position distributions are double-peaked, and these two peaks are 
separating when the energy of the hit increases.
Such histograms were used to determine the dependence of 
the distance $|Z_{hit}-Z_{v}|$ 
on the energy loss $\Delta E_{norm}$ shown in Fig.~4.
The hits from particles registered even at 100~cm from the vertex 
can be used to estimate the vertex position. 
The uncertainty of this energy loss based extrapolation 
from a single hit is always smaller than the one  
obtained from the simple geometrical arguments described at the beginning of this 
section, and for large fraction of hits it is less than 3~cm.  
The main contribution to this
error is the effect of Landau fluctuations on the energy loss
measurements.

\begin{figure}[bt]
\includegraphics[width=7.5cm]{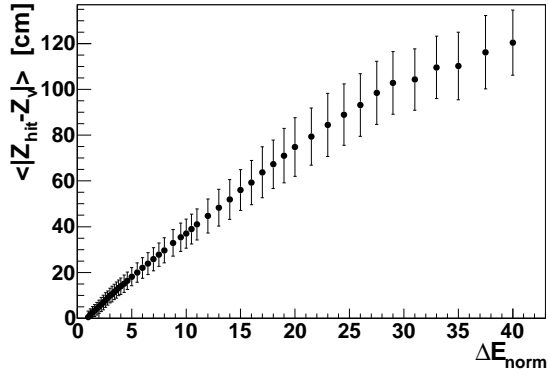}

\caption{ \label{Fig4}
The mean distance of the hit from the vertex, $|Z_{hit}-Z_{v}|$, as a function of 
the energy loss, $\Delta E_{norm}$. The error bars represent the values of 
the width, $\sigma(|Z_{hit}-Z_{v}|)$, of the distributions
of $|Z_{hit}-Z_{v}|$ distances obtained for fixed values of $\Delta E_{norm}$.
}
\end{figure}

In the analysis of the real data
we first reconstruct the hits and their energy loss from measured signals.
For rejection of noise we apply a threshold of 0.4~$\Delta E_{MIP}$,
i.e. up to about three times the typical noise level.
A signal in a single, isolated pad is then accepted as a hit when
it exceeds 0.6~$\Delta E_{MIP}$, while
neighboring pads with signals are merged along the beam direction into
a multi-pads hit.
Each multi-pad hit has an energy larger than 0.6~$\Delta E_{MIP}$,
but hits with three or more pads are tested more precisely.
A primary particle producing 
such a hit traverses the whole width of the middle pads (0.27~cm) and 
leaves there a signal much larger than $\Delta E_{MIP}$. 
It is thus required that at least 40\% 
of the expected mean energy loss is registered in each of the middle pads of the hit.

\section{Calculation of the vertex position}
\label{Sec4}

In the first step of the vertex reconstruction procedure
we find the hits and calculate their energy loss signals. 
Due to the presence of background hits, a simple weighted 
mean of vertex position estimates will not be accurate, even 
if the sign of the distance to the vertex is correctly guessed for each hit.
More appropriate is a maximum likelihood method using a probability
function of one variable, $z$. We have implemented two methods using different functions.
The values of these functions are calculated for many points along the beam line 
to find the most likely vertex position.

In the first method, at each tested vertex position 
the energy loss registered in the silicon is for all hits rescaled 
to the energy loss expected for a particle traversing the thickness of the sensor. 
The hit is accepted when this energy is between 
0.8~$\Delta E_{MIP}$ and 1.25~$\Delta E_{MIP}$
and the number of accepted hits is counted. 
Neglecting background hits, this function at the vertex should be equal to the 
number of primary particles with hits. 

In the second method, histograms of $Z_{hit}-Z_{v}$, obtained 
for fixed values of energy loss $\Delta E_{norm}$
(like these shown in Fig.~3), define appropriately 
normalized and parameterized probability density functions 
$P(\Delta E_{norm}, Z_{hit}-Z_{v})$.  We are using
40 functions of $Z_{hit}-Z_{v}$, obtained for 40 different $dE$ values.
In order to speed up calculations, they are approximated piecewize by linear segments. 
Parameters of these approximations 
are interpolated for intermediate $dE$ values.
Then, at each tested hypothetical vertex position, 
a product, $\Pi P$, of the values of this probability density 
function for all hits is calculated. 
The hits with very small probabilities $P$ (or with $P$=0) are usually due to 
background particles. While factors of zero should be avoided 
in the product of probabilities, it is important not to 
neglect the background in the calculation of $\Pi P$. 
In order to treat all background hits in a consistent way, 
all small $P$ values are rounded up to $P_{min}$ (about 5\% of the maximal probability).
In tests with smaller $P_{min}$ values, vertex reconstruction 
results were very similar. The product $\Pi P$ contains factors for
all hits: $P_{min}$ values for background hits and larger (but not identical) values
for hits regarded to be from primaries. Usually $P_{min}$ values appear 
many times at all considered $z$ positions, thus in our calculations 
we neglect the same, largest common number of them. 
In this way we reduce the time consuming 
operations and avoid numerical problems with multiplication of 
a large number of very small values (underflow).

\begin{figure}[tb]
\includegraphics[width=7.5cm]{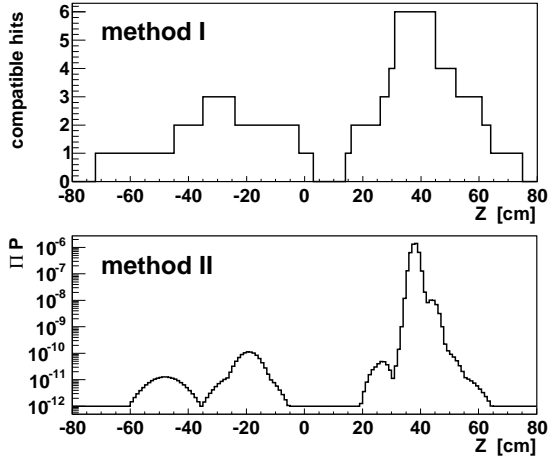}

\caption{ \label{Fig5}
Examples of the histograms used to determine the position of the vertex 
in two versions of the algorithm described in the text, obtained for
the vertex position at $Z_{v}$=40~cm and 6 primary tracks with hits. Secondary 
particles production and detector effects are not included.
}
\end{figure}
In both methods, the calculated values of probability function for
each hypothetical vertex position are collected in histograms
similar to those shown in Fig.~5.
The maximum of each histogram is close 
to the actual position of the vertex. When the true vertex 
is far from $z$=0, there is usually a second, wider maximum 
at the opposite $z$ side, such as those in Fig. 5 at $z$=$-$30~cm (top histogram)
or $z$=$-$20~cm (bottom histogram), 
due to grouping of the estimates
of the vertex position at the wrong side of the hit. Finding the vertex position
is more difficult for events from either real data, or from simulations
that include secondary particle generation. 
The hits from background can enhance the second maximum or produce
another one while the main maximum becomes wider and lower
because of the noise. Therefore
we must not accept the position of the maximum as a reliable vertex, when 
it comes from very few hits or when there is a second maximum of similar height.
The selection of the cuts is a compromise between efficiency and purity of the
accepted vertices.
The second method described above was found to be more efficient and accurate,
and only the performance of this method is presented in the next sections.
 
\begin{figure}[tb]
\includegraphics[width=7.5cm]{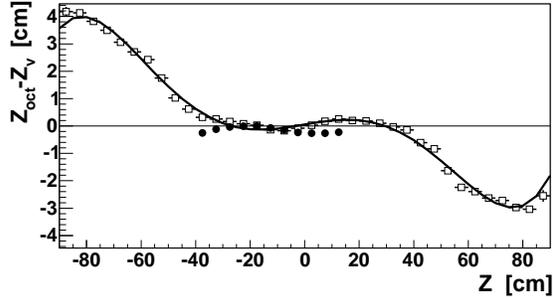}

\caption{ \label{Fig6}
The correction for the bias of $octagon$ vertex reconstruction method.
The uncorrected difference, $\langle Z_{oct,raw}-Z_{v} \rangle$ (open squares), 
as a function of the vertex position is fitted
by a polynomial (curve).
For real data the precise vertex obtained 
from tracks reconstructed in the PHOBOS spectrometer, $Z_{spec}$,
substitutes the true vertex position in the calculations
of similar difference, $\langle Z_{oct}-Z_{spec}\rangle$ (full circles), 
in which already corrected value of $Z_{oct}$ is used.
The simulations and the data for
Au+Au collisions at $\sqrt{s_{_{NN}}}$=200~GeV were used.
}
\end{figure}

\section{Efficiency and accuracy of the vertex reconstruction algorithm}
\label{Sec5}

The properties of the vertex finding algorithm were studied using 
Monte Carlo simulations, based on the GEANT \cite{geant} package, with
production of secondary particles and including all detector effects 
(geometrical acceptance, electronics noise, digitization and calibration).
The collisions of Au+Au, Cu+Cu, d+Au and p+p at energies measured with the PHOBOS
detector were simulated
with the vertex positions over a range of  $\pm$120~cm,
which substantially exceeds
the span of the $octagon$ ($\pm$55~cm).

The $octagon$ vertex reconstruction algorithm relies on the parameterized 
probability function $P(\Delta E_{norm}, Z_{hit}-Z_{v})$ extracted from 
simulations of the response from primary particles. These are derived from energy
loss values not affected by secondary processes or detector noise, thus 
a bias in the estimate of the vertex position, $Z_{oct}$, is very probable.
Fig.~6 shows that $Z_{oct}$ is systematically shifted towards the center
of the $octagon$ detector, in the direction where more hits are registered.
Hits from secondary particles, which usually traverse silicon sensors 
at angles larger than the primaries (in the same part of the detector), 
tend to attract the reconstructed vertex. Also for the hits from 
these primary particles, which traverse 2 pads, the signal in one pad 
may be lost when the pad is defective or may be rejected as noise if it is smaller
than 0.4~$\Delta E_{MIP}$. These effects result in an underestimate of the distance 
of the hit from the vertex. It is also possible,
that the $P(\Delta E_{norm}, Z_{hit}-Z_{v})$ parameterization 
is responsible for part of the bias.
The observed bias is relatively small in the most useful 
range $|Z_{oct}|$\textless 40~cm and is very similar for events with different 
multiplicities. Rather than trying to remove it by modifications 
of the algorithm, we apply a universal correction using the curve from Fig.~6.
This bias correction is tested in Fig.~6 for real events by comparing the
$octagon$ vertex, $Z_{oct}$, with the vertex precisely determined 
using tracks found in the spectrometer, $Z_{spec}$. 
We observe a small systematic difference between the true and reconstructed vertex
position which will be discussed later.

\begin{figure}[tb]
\includegraphics[width=7.5cm]{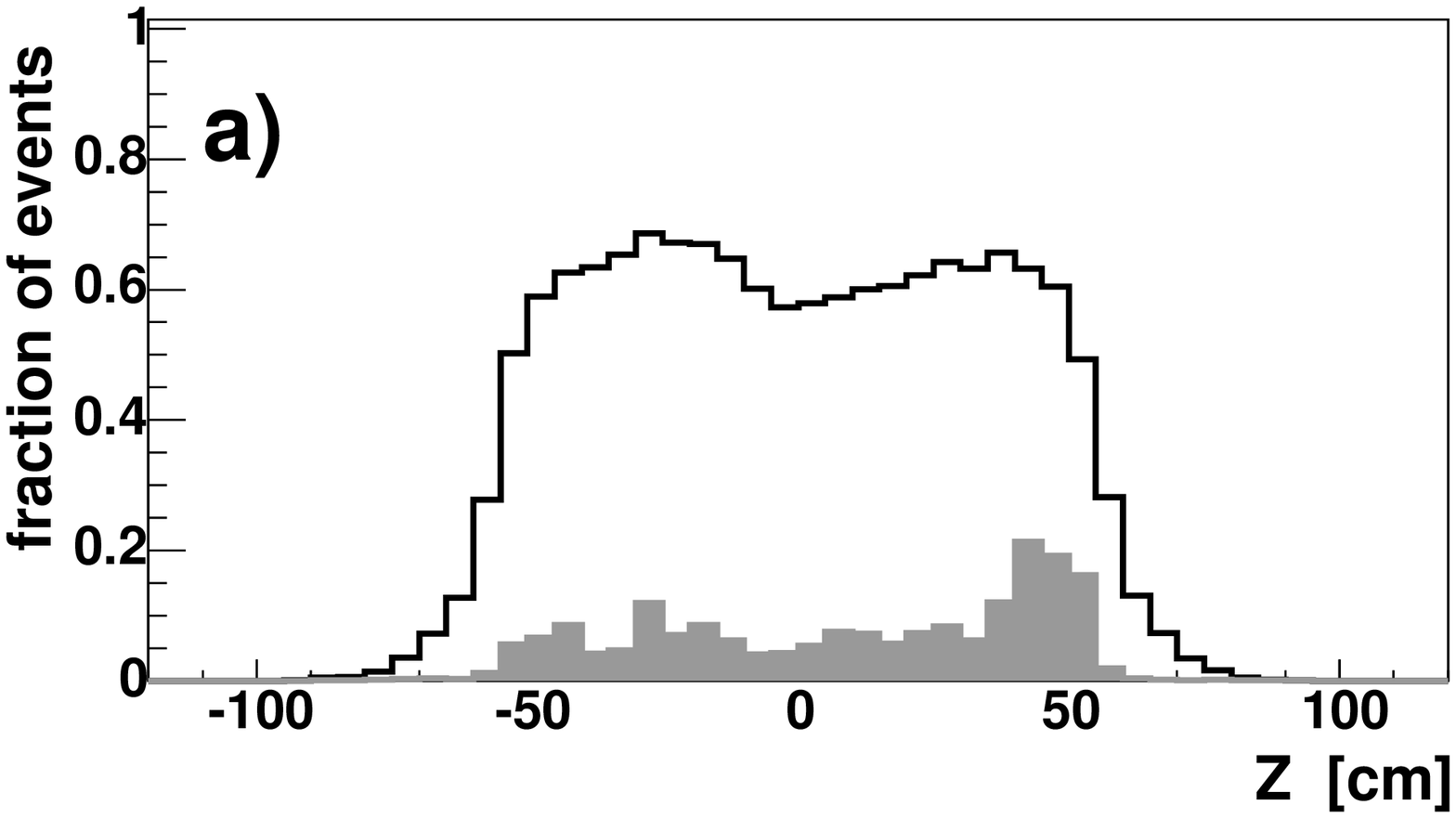}
\includegraphics[width=7.5cm]{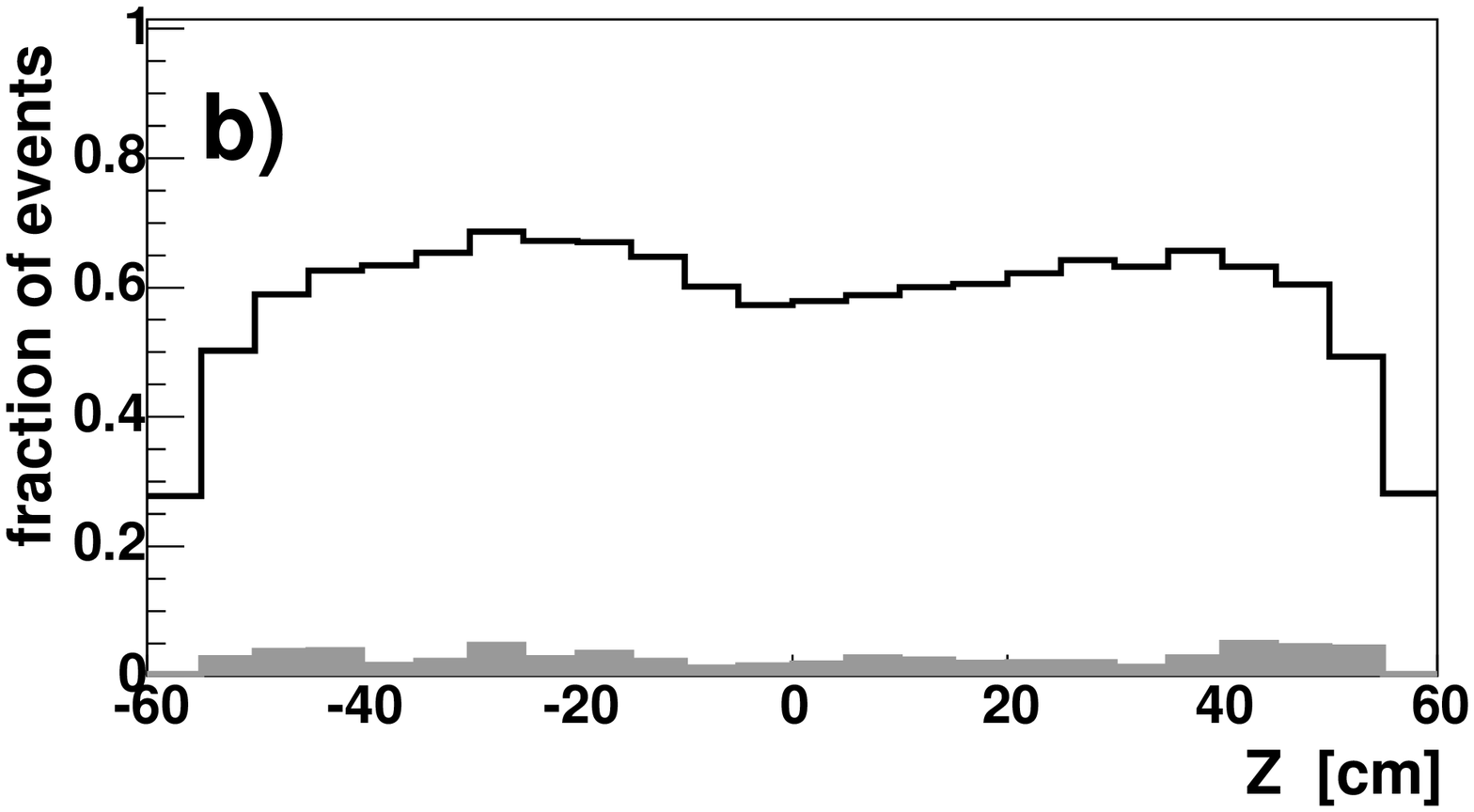}

\caption{ \label{Fig7}
The efficiency of the vertex reconstruction (histograms) in p+p collisions at
$\sqrt{s_{_{NN}}}$=200 GeV as a function of the vertex position,
$Z_{v}$, for true vertex position limited to $\pm$120~cm (a) and $\pm$60~cm (b).
The accepted events have $|Z_{oct}-Z_{v}|$\textless 5~cm. The shaded areas show
the distribution of $Z_{oct}$ positions for remaining events, in which the 
vertex was reconstructed, but in a wrong position.
}
\end{figure}

In the analysis of the vertex reconstruction efficiency we start 
with the most difficult case - the sample of p+p collisions which
has the smallest mean multiplicity of produced particles. 
The $octagon$ vertex reconstruction algorithm 
finds the vertices in about 60\% of events in the $z$ range 
$\pm$60~cm  (Fig.~7).
For some of the events with the real vertex outside this acceptance range 
a false vertex is found, especially at positive $z$. There are more 
background hits registered, mostly from secondary particles produced 
in the coils and the yoke of the PHOBOS magnet \cite{detector}.
Such events can be rejected by 
restricting the vertex range using information from the trigger counters.
If the true vertex range is limited to $\pm$60~cm the fraction
of additional events with wrong vertex drops below 3\% (Fig.~7b).

\begin{figure}[tb]
\includegraphics[width=7.5cm]{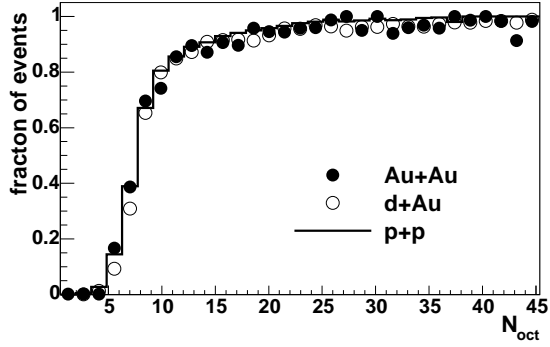}

\caption{ \label{Fig8}
The efficiency of the vertex reconstruction as a function of the number 
of charged primary particles registered in the octagonal multiplicity 
detector, $N_{oct}$, for Au+Au, d+Au and p+p collisions at $\sqrt{s_{_{NN}}}$=200 GeV.
}
\end{figure}

The vertex reconstruction performance depends on the 
event multiplicity and a pertinent parameter is the number 
of charged primary particles leaving hits in the $octagon$, $N_{oct}$.
It is equal to about 50\% of the total number of charged primaries.
In Fig.~8 we can see that the algorithm is over 95\% efficient
if $N_{oct}$ is greater than 15 
and the efficiency drops with decreasing multiplicity down to zero
for $N_{oct}$\textless~4.
This dependence is very similar for all types of
collisions. A slight difference can be noticed only for d+Au collisions, 
which have non-symmetric rapidity distribution.

\begin{figure}[tb]
\includegraphics[width=7.5cm]{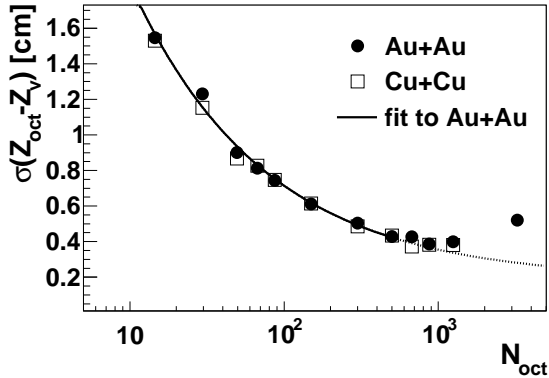}

\caption{ \label{Fig9}
The error of the reconstructed vertex position, $\sigma(Z_{oct}-Z_{v})$,  
as a function  of the number of charged primary particles registered 
in the $octagon$, $N_{oct}$, for Au+Au and Cu+Cu collisions 
at $\sqrt{s_{_{NN}}}$=200 GeV. 
}
\end{figure}

The error of the reconstructed vertex position,
presented for two types of collisions in Fig.~9, also depends on event multiplicity.
It is fitted for \mbox{$N_{oct}$\textless 500} and then extrapolated  
as \mbox{$\sigma(Z_{oct}-Z_{v})$~=}~0.19~+~\mbox{5.26\,/\,$\sqrt{N_{oct}}$.}
The value of the second parameter of the fit, 5.26, is consistent 
with $\sigma(Z_{hit}-Z_{v})$ obtained for single hits at fixed energy loss (Fig.~4).
For the highest multiplicities, \mbox{$N_{oct}$\textgreater 1000,} the vertex position error
deviates from the extrapolation of the fit and even increases. 
This effect is due to high detector occupancy, as at large
multiplicities the chance that two particles hit the same or neighbor pads increases.
In such cases, a common hit with twice the energy loss is formed and  
the algorithm incorrectly estimates the distance of this hit from the vertex. 
Fortunately, for events with such large multiplicities we are able 
to reconstruct the vertex using more precise, track based methods \cite{time05}.

\begin{table}[tb]
\begin{tabular}{|l|c|c|c|c|c|}
\hline
~Colliding~ & ~$\sqrt{s_{_{NN}}}$~ &  ~$\langle N_{oct}\rangle$~  
                        & Correct  &  Wrong   &  Error   \\
~System &  [GeV]  &     & vertex  &  vertex   &  ~$\sigma(Z_{oct}-Z_{v})$~  \\
\hline
Au+Au & 200      & 619 &  94.5\%       &  0.4\%  &   0.58~cm  \\
Au+Au & ~19      & 252 &  91.8\%       &  0.4\%  &   0.69~cm  \\
Cu+Cu & 200      & 198 &  90.5\%       &  0.5\%  &   0.64~cm  \\
d+Au  & 200      & ~39 &  87.6\%       &  0.7\%  &   1.03~cm  \\
p+p   & 200      & ~~9 &  58.6\%       &  2.6\%  &   1.63~cm  \\
\hline
\end{tabular}
\vspace{0.1cm}
\caption{
\label{Tab1}
Summary of the properties of the vertex reconstruction algorithm: 
the fraction of events with correctly reconstructed vertex 
(i.e. $|Z_{oct}-Z_{v}|$\textless 5~cm), the fraction of events with wrong vertex 
($|Z_{oct}-Z_{v}|\geqslant$5~cm)
and the RMS of a Gaussian fit to the $Z_{oct}-Z_{v}$ distribution.
The true vertex range is limited to $|Z_{v}|$\textless 60~cm. 
}
\end{table}

The performance of the vertex reconstruction algorithm for several types of collisions 
measured by the PHOBOS experiment is summarized in Table~1. 
For nucleus-nucleus collisions, the vertex finding  efficiency 
is close to 90\% and a small (\textless 1\%)
admixture of events with wrongly reconstructed vertex is accepted. For p+p collisions
this admixture grows to 2.6\% while the efficiency drops below 60\%.
The efficiency of precise methods using tracks found in the spectrometer or the vertex detector, 
is always smaller, especially for the p+p collisions 
(1.4\% and 6.4\% respectively), even with the evaluation restricted to  
narrower vertex range $|Z_{v}|$\textless 10~cm, 
in which these methods should perform best \cite{time05}.
The average vertex position errors listed in Table~I were obtained for 
full samples of events with reconstructed vertex. 
They reflect a convolution of the multiplicity 
distribution in the collisions and the dependence of the error on multiplicity. 
Interestingly, in the analysis of special simulations without production
of secondary particles we find very similar efficiency and accuracy
when the same reconstruction procedure is applied.
The efficiency for the events without background is even smaller, probably 
because quite often secondaries following the direction of their parents 
help to increase the main maximum, which otherwise would not be 
accepted by the quality cuts.
The similarity of results for simulations with and without secondary particle
production proves that with the present cuts the reconstruction is not very sensitive
to the background.

\section{Vertex reconstruction for real data}
\label{Sec6}

The accuracy of the reconstructed $octagon$ vertex can be studied not only
using Monte Carlo simulations, but also with the real data. 
In this case we use events in which the vertex was also 
reconstructed by other methods, for example based
on information from the spectrometer, $Z_{spec}$. The vertex position is then determined
with much better accuracy, however only for events with large multiplicities 
(about 50\% of Au+Au collisions at $\sqrt{s_{_{NN}}}$=200 GeV). For these events 
the value of $\langle Z_{oct}-Z_{spec}\rangle$, in a wide range of vertex positions is 
shown in Fig.~6. There is a systematic difference (about $-$0.2~cm), which is not 
removed by the bias correction. It may be due to a weak multiplicity dependence
that is neglected in the bias correction, misalignments of the $octagon$ sensors
or a difference between the energy loss 
distributions in the simulations and in the data.
The last possibility is most probable, as also the width, $\sigma(Z_{oct}-Z_{spec})$,
of the $Z_{oct}-Z_{spec}$ distribution is smaller for the real data
than for the simulations (about 0.36~cm and 0.45~cm, respectively). 
A longer tail of the energy loss distribution in the simulations 
can lead to these differences.

The residual bias observed in the data may be additionally corrected 
for in the analysis.
However, in the low multiplicity events for which the $octagon$ vertex 
is most useful, such a small correction is hardly relevant, as in this
case the vertex reconstruction error is larger than 1~cm.

\section{Summary}

In this paper we present a novel method of vertex reconstruction using
a large acceptance detector with a single layer of silicon sensors. 
From all algorithms developed for the PHOBOS experiment the $octagon$ method
allows to reconstruct vertices in the widest $z$-range and with by far 
the largest efficiency, however with
a significant error of 0.5-2~cm, depending on event's multiplicity. 
The more precise algorithms using the information from the spectrometer or 
the vertex detector are less efficient and in the events with
low multiplicity may find a false vertex position. The $octagon$ vertex 
enables verification of vertices given by other methods or substitutes them when 
other methods fail. It is extensively used in studies 
of d+Au \cite{dau1,dau2,dau3,dau4,dau5} and p+p \cite{pp} collisions.

\noindent{\bf Acknowledgments}
We thank our colleagues, Birger Back and Constantin Loizides, for their help 
during preparation of the manuscript. 
This work was partially supported by U.S. DOE grants
%
%
DE-AC02-98CH10886,
DE-FG02-93ER40802,
DE-FC02-94ER40818,  
DE-FG02-94ER40865,
DE-FG02-99ER41099, and
W-31-109-ENG-38, by U.S.
NSF grants 9603486, 
0072204,            
and 0245011,        
by Polish KBN grant 1-P03B-062-27(2004-2007),
by NSC of Taiwan Contract NSC 89-2112-M-008-024, and
by Hungarian OTKA grant (F 049823).

\end{document}